\documentclass[aps,pra,reprint,showpacs,amsmath,amssymb,groupedaddress,floatfix,nofootinbib,superscriptaddress] {revtex4-1}

\usepackage{bm}    
\usepackage{bbm}
\usepackage{amssymb}
\usepackage{color}
\usepackage{nicefrac}
\usepackage{graphicx}
\usepackage[utf8]{inputenc}
\usepackage[english]{babel}
\usepackage{todonotes}
\usepackage{ulem}
\usepackage{amsfonts}
\usepackage[T1]{fontenc}

\newcommand{\e}{\mathrm{e}}
\newcommand{\erf}{\mathrm{erf}}

\newcommand{\psidouble}{\Psi^{double}}
\renewcommand{\(}{\left(}
\renewcommand{\)}{\right)}
\newcommand{\rspdm}{\rho}

\begin{document}
\title{Quantum dynamics of two trapped bosons following infinite interaction quenches}
\author{L. M. A. Kehrberger}
\email{lkehrber@physnet.uni-hamburg.de}
\affiliation{Zentrum f\"ur Optische Quantentechnologien, Universit\"at Hamburg, Luruper Chaussee 149, 22761 Hamburg, Germany}

\author{V. J. Bolsinger} 
\email{vbolsing@physnet.uni-hamburg.de}
\affiliation{Zentrum f\"ur Optische Quantentechnologien, Universit\"at Hamburg, Luruper Chaussee 149, 22761 Hamburg, Germany}
\affiliation{The Hamburg Centre for Ultrafast Imaging, Universit\"at Hamburg, Luruper Chaussee 149, 22761 Hamburg, Germany}

\author{P. Schmelcher}  
\email{pschmelc@physnet.uni-hamburg.de}
\affiliation{Zentrum f\"ur Optische Quantentechnologien, Universit\"at Hamburg, Luruper Chaussee 149, 22761 Hamburg, Germany}
\affiliation{The Hamburg Centre for Ultrafast Imaging, Universit\"at Hamburg, Luruper Chaussee 149, 22761 Hamburg, Germany}

\date{\today}

\begin{abstract}
We investigate the quantum dynamics of two identical bosons in a one-dimensional harmonic trap following an interaction quench from zero to infinite interaction strength and vice versa.
For both quench scenarios, closed analytical expressions for the temporal evolution of the wave function as well as the Loschmidt Echo are found and the dynamics of the momentum distribution as well as the reduced single-particle density matrix are analyzed.
We observe a crossover of these quantities between bosonic, "symmetrized" fermionic and fermionic properties.
Furthermore, several combined quenches are analyzed as well.
\end{abstract}

\maketitle

\section{Introduction}

The non-equilibrium quantum dynamics of ultracold bosonic systems has become a field of intense research over the past two decades \cite{Morsch2006,Bloch2008,Polkovnikov2011}.
Most theoretical works on ultra-cold bosonic ensembles are of numerical nature, since exactly solvable models are rare.
However, if an analytical solution is found, it can help to gain intuition and a deeper understanding of the temporal evolution of the underlying quantum system, to validate numerical simulations as well as sophisticated approximations.
Also analytic result can be used as a starting point for numerical methods as well as perturbative treatments.

When analytically approaching many-body systems, one of the main challenges is to handle the inter-particle interaction potential properly. 
If the inter-particle distance is much larger than the interaction range, the interaction potential can be approximated by a two-particle contact interaction \cite{Huang1957,Derevianko2005}.
Among the systems solved analytically within this approximation are bosonic ensembles in one spatial dimension with periodic boundary conditions \cite{Lieb1963,Lieb1963b}, or two bosons trapped in isotropic or anisotropic harmonic potentials \cite{Busch1998,Idziaszek2006,Cirone2001}.
Furthermore, for one-dimensional problems, exact solutions can be obtained in the unitary regime or Tonks-Girardeau via the application of the Fermi-Bose mapping \cite{Tonks1936,Girardeau1960}, i.e., the Jordan-Wigner transformation, which links impenetrable bosons to non-interacting fermions.

Experimentally, the Tonks-Girardeau regime can be achieved by modifying the transversal trap frequencies \cite{Olshanii1998} or by exploiting  Feshbach resonances \cite{Chin2010}.
With the possibility of experimentally tuning the interaction strength to very high values, the strongly interacting bosonic regime is of current interest \cite{Chevy2016} and the corresponding density distribution \cite{Kinoshita2004}, the momentum distribution \cite{Paredes2004}, correlations  \cite{Kinoshita2005}, collective modes \cite{Haller2009}, transport properties \cite{Palzer2009} as well as fluctuations \cite{Jacqmin2011} of Tonks-Girardeau gases have been studied.
Theoretical investigations regard the strong interacting regimes include the tunneling dynamics \cite{Zollner2008a,Zollner2008b}, ground-state fragmentation \cite{Zollner2006, Cederbaum2005}, or quench dynamics from the non-interacting to the unitary regime for different trap geometries and dimensions \cite{GarciaMarch2016, Mazza2014, Sykes2014}.
Here, properties such as the density-density correlation function \cite{Marton2014,Bastianello2017}, breathing oscillations \cite{Fang2014},  momentum distribution dynamics \cite{Atas2016}, or the quantum entanglement between two bosons have been studied \cite{Mack2002,Sun2006,Goold2009,Sowinski2010}.

In this work, we analytically derive a closed expression for the time-dependent wave function (and its Loschmidt Echo) of two interacting bosons in a one-dimensional harmonic trap for both an interaction quench from zero to infinity and vice versa, i.e., from infinity to zero.
As initial states, we choose various eigenstates of the initial Hamiltonian.
For the evolution of the respective ground states, we calculate the time evolution of the reduced single-particle density matrices as well as the momentum distributions.
Furthermore, we discuss multiple quench scenarios, where we consecutively quench the interaction strength from zero to infinity and then back to zero as well as from infinity to zero and back to infinity.

This work is structured as follows:
In section \ref{sec:Stationary}, we give a brief sketch of the upcoming calculations and review the eigenvalues and eigenfunctions of both the non-interacting and the infinitely-strong interacting Hamiltonians.
In section \ref{sec:toInfinity}, we study the dynamics of the interaction quench from zero to infinity, and in section \ref{sec:fromInfinity}, the reversed quench (from infinity to zero) is analyzed and a mathematical connection between these two quench scenarios is shown.
For both quench scenarios and the respective ground states as the initial state, we derive a closed expression for the temporal evolution of the wave function.
Using the previous results, we study different combinations of those two interactions quenches in section \ref{sec:double}.
Finally, a conclusion of our findings is given in section \ref{sec:conclusion}.

\section{Hamiltonian and its eigenfunctions} \label{sec:Stationary}

In this section, we set the necessary theoretical groundwork in order to perform interaction quenches of two identical bosons between the non-interacting and the Tonks-Girardeau regime.
First, we describe the setup by writing down the underlying Hamiltonians for both regimes and review the corresponding eigenfunctions and eigenvalues.
Second, we give the definitions of the reduced single-particle density matrix (SDM) as well as the momentum distribution and calculate them for the  ground states in the respective regimes.
Last, the framework for calculating the temporal evolution of a wave function is sketched, and we define the fidelity as well as the Loschmidt echo, quantities used to characterize the dynamics.

\subsection{Stationary solution}

Our setup consists of two identical bosons at positions $z_1$ and $z_2$ in a one-dimensional, harmonic potential which interact via a $\delta$-potential with interaction strength $\kappa$.
The Hamiltonian in harmonic oscillator units ($m=\hbar=\omega=1$) reads
\begin{align}
&\hat{H}^{\kappa}=-\frac{1}{2}(\partial_1^2+\partial_2^2)+\frac{1}{2}(z_1^2+z_2^2)+\sqrt{2}\kappa \delta(z_1-z_2),	 \nonumber
\end{align}
where we have added the factor $\sqrt{2}$ for convenience.
The Hamiltonian can be separated into scaled center-of-mass (CM) and relative (rel) coordinates, $Z=(z_1+z_2)/\sqrt{2}$ and $z=(z_1-z_2)/\sqrt{2}$, respectively:
\begin{align}
&\hat{H}^{\kappa}=\underbrace{-\frac{1}{2}\partial_Z^2+\frac{1}{2}Z^2}_{\hat{H}_{\mathrm{CM}}}
\underbrace{-\frac{1}{2}\partial_z^2+\frac{1}{2}z^2+\kappa\delta(z)}_{\hat{H}^{\kappa}_{\mathrm{rel}}}. \nonumber
\end{align}
The CM coordinate is not affected by the interaction potential, and we label the solution of the CM-Hamiltonian by $\chi_n(Z)$, which are harmonic oscillator functions.
Thus, we have to deal only with the relative Hamiltonian, which is an effective single-particle problem of one particle in a harmonic trap with a delta-potential at the origin.

For the non-interacting case $\kappa=0$, the relative Hamiltonian eigenfunctions are simply those of a harmonic oscillator,
\begin{align}
\psi_n(z) & = b_n H_n(z)\e^{-\frac{z^2}{2}},                \nonumber\\
      b_n & = \frac{1}{\pi^{\frac 1 4}\sqrt{ 2^n n!}},		\label{psin}
\end{align}
where $H_n$ are the physicist's Hermite polynomials with $n \in \mathbb{N}_0$.
The corresponding energy-eigenvalues are given by $E_n=n+\frac{1}{2}$.
Since the particle exchange symmetry is reflected in the parity symmetry of the relative wave functions, we call quantities based on the wave functions $\psi_{2n}(z)$ \textit{bosonic}\footnote{Recalling the relation $H_n(-z)=(-1)^nH_n(z)$}.

For infinitely strong repulsive interaction $\kappa=\infty$, we obtain for the relative Hamiltonian eigenfunctions \cite{Busch1}:
\begin{align}
\phi_{2n}(z)=\psi_{2n+1}(|z|), \nonumber\\ 
\phi_{2n+1}(z)=\psi_{2n+1}(z) \nonumber
\end{align}
with doubly-degenerate energy eigenvalues $\epsilon_{2n}=\epsilon_{2n+1} = E_{2n+1}$.
We refer to properties resulting from the wave functions $\phi_{2n+1}(z)$ as  \textit{fermionic} and from $\phi_{2n}(z)$ as \textit{symmetrized fermionic}, since the $\phi_{2n}(z)$ have the probability density of the fermionic wave functions but bosonic symmetry.

\subsection{Density matrix and momentum distribution}

For an arbitrary two-particle wave function $\Xi(z_1,z_2)$, the reduced single-particle density matrix (SDM) $\rspdm(z_1,z_1')$, which characterize the coherence between $z_1$ and $z_1'$, is defined by
\begin{align}
\rspdm(z_1,z_1'):=&\int_{-\infty}^\infty  \Xi(z_1,z_2) \Xi^\ast(z_1',z_2)  \,dz_2 , 			\label{SDM}
\end{align}
from which the reduced single-particle density is given as the diagonal $\rho(z_1):=\rspdm(z_1,z_1)$.
The SDM can be decomposed into eigenvectors (natural orbitals) $\beta_i$, 
\begin{align}
\rspdm(z_1,z_1') = \sum_{i} \lambda_i \beta_i(z_1) \beta_i^*(z_1'), \nonumber
\end{align}
with the positive eigenvalues (natural populations) $\lambda_i$.
From the SDM, we can obtain the momentum distribution $n(k)$ via 
\begin{align}
n(k):=\frac{1}{2\pi}\int_{-\infty}^\infty \int_{-\infty}^\infty \rho(z_1,z_1')\e^{-ik(z_1-z_1')}\,dz_1\,dz_1', 		\label{nvonk}
\end{align}
or better numerically accessible, via the Fourier transform $\tilde{\beta}_i(k)$ of the natural orbitals
\begin{equation}
n(k)=\sum_i\lambda_i|\tilde{ \beta_i}(k)|^2. \nonumber
\end{equation}
The SDMs and the momentum distributions based on the bosonic (b) and fermionic (f) relative Hamiltonian  ground states $\psi_0$ and $\phi_1$, respectively, can be easily calculated [assuming $\chi_0(Z)$ for the CM motion]:
\begin{align}
\rspdm_b(z_1,z_1')&=\frac 1 {\sqrt{\pi}} \e^{-\frac 1 2 (z_1^2+z_1^{\prime 2})}, \nonumber \\
\rspdm_{f}(z_1,z_1')&=\frac{1+2z_1z_1'}{2{\sqrt{\pi}} }\e^{-\frac 1 2 (z_1^2+z_1^{\prime 2})} \nonumber
\end{align}
from which we obtain the normalized momentum distributions:
\begin{align}
n_b(k)=&\frac{1}{\sqrt{\pi}} \e^{-k^2}, \nonumber \\
n_f(k)=&\frac{1+2k^2}{2 \sqrt{\pi}} \e^{- k^2}. \nonumber
\end{align}
The case of the symmetrized fermionic (sf) ground state, $\phi_0(z)=\psi_1(|z|)$, is more challenging.
In order to calculate the SDM, we split the integral up assuming $z_1'>z_1$ and keep generality by including the sign of $z_1'-z_1$ in front of the integral\footnote{We choose the error function's definition with the pre-factor $2/\sqrt[]{\pi}$, i.e. $\erf(z):=\frac{2}{\sqrt[]{\pi}}\int_0^z\e^{-x^2}\,dx$ }
\begin{align}
&\rspdm_{sf}(z_1,z_1') \nonumber \\ 
&=\frac{\e^{-\frac 1 2 (z_1^2+z_1^{\prime 2})}}{\pi} \int_{-\infty}^\infty 		|z_1-z_2||z_1'-z_2|\e^{-\frac{z_1^2}{2}-\frac{z_1^{\prime 2}}{2}-z_2^2} \, dz_2		\nonumber\\
&=	\frac{\e^{-\frac 1 2 (z_1^2+z_1^{\prime 2})}}{\pi}\int_{-\infty}^\infty (z_1-z_2)(z_1'-z_2)\e^{-z_2^2}\,dz_2 		\nonumber\\
&-2\frac{\e^{-\frac 1 2 (z_1^2+z_1^{\prime 2})}}{\pi}\mathrm{sgn(z_1'-z_1)}\int_{z_1}^{z_1'} (z_1-z_2)(z_1'-z_2)\e^{-z_2^2}\,dz_2		\nonumber\\
&=\rspdm_f(z_1,z_1')+\frac{\e^{-\frac{z_1^2}{2}-\frac{z_1^{\prime 2}}{2}}}{\pi}\mathrm{sgn(z_1'-z_1)}		\nonumber\\ 
&\left(z_1'\e^{-z_1^2} - z_1 \e^{-z_1'^2}+\sqrt{\pi } \left( z_1 z_1'+\frac 1 2\right) (\text{erf}(z_1)-\text{erf}(z_1'))\right). \nonumber
\end{align}
Its momentum distribution $n_{sf}(k)$ has to be determined numerically (see Sec.\  \ref{sec:toInfinity}).

\subsection{Temporal evolution and fidelity}
%-------------------------------------------

Here, we give a quick reminder how the temporal evolution of a general, initial wave function  $|\Psi(t_0)\rangle$ can be obtained when the complete set of stationary eigenfunctions $|\xi \rangle$ with eigenvalues $E_\xi$ of the Hamiltonian $\hat{H}$ is known.
Invoking the Schr\"odinger equation  $i\partial_t|\Psi(t)\rangle=\hat{H}|\Psi(t)\rangle$, we can make the ansatz 
\begin{align}
|\Psi(t)\rangle =  \mathrm{e}^{-i\hat{H}(t-t_0)}|\Psi(t_0)\rangle 
=\sum_{\xi=0}^\infty\mathrm{e}^{-i E_\xi (t-t_0)} |\xi \rangle \langle \xi |\Psi(t_0)\rangle,
\label{howtoevolve}
\end{align}
In this manner, the solution of the Schr\"odinger equation can be reduced to the calculation of the overlap integrals $\langle \xi|\Psi(t_0)\rangle$ and a summation of all eigenfunctions weighted by the overlap integrals with a time-dependent phase factor.

The sensitivity of the temporal evolution of an initial state to perturbations can be judged by the overlap between the initial state and its temporal evolution, the auto-correlation $L(t) := \langle\Psi(t_0)|\Psi(t)\rangle$, which is related to the Loschmidt echo $\mathcal{L}(t)$ via $\mathcal{L}(t):=|L(t)|^2$ and to the fidelity between the two wave functions via $F(|\Psi(t_0)\rangle,|\Psi(t)\rangle):=|L(t)|$.
The initial wave function is completely recovered for $L(t) = 1$.
However, if $L(t)=0$, then the time evolved state becomes orthogonal to the initial state.

\section{Interaction quench from zero to infinity} \label{sec:toInfinity}

In the first part of this section, we perform an interaction quench (at $t_0=0$) from zero to infinity, $\kappa=0\to\infty$, of various, initial eigenstates of the non-interacting regime.
We analytically derive the temporal evolution of these initial states as well as the overlap between the time-evolved and initial states.
In the second part, we focus on the time evolution of the ground state for which we derive a simple closed expression, and study its SDM as well as its momentum distribution.

\subsection{Quench of arbitrary eigenstates}

Following Eq.\ \eqref{howtoevolve}, we solve the Schr\"odinger equation $\hat{H}^{\kappa=\infty}_{\mathrm{rel}}|\Psi_m(t)\rangle=i\partial_t|\Psi_m(t)\rangle$ with initial conditions $|\Psi_m(t=0)\rangle = | \psi_{2m}\rangle $ \footnote{Since the odd eigenstates $\psi_{2m+1}$ are eigenfunctions in both regimes they lead to a trivial temporal evolution and are neglected in the following discussion.} such that the time evolution of the wave function reads  $|\Psi_m(t) \rangle = \sum_{n=0} \exp(-i\epsilon_{2n}) c_{mn} | \phi_{2n} \rangle$. 
Therefore, we need to calculate the overlap coefficients $c_{mn} \equiv \langle\phi_{2n}|\psi_{2m}\rangle$\footnote{Mind that $\langle\phi_{2n+1}|\psi_{2m}\rangle$=0} between the symmetrized fermionic and the bosonic wave functions, given by the integral:
\begin{align}
 c_{mn}=& 2 b_{2m} b_{2n+1} \underbrace{\int_{0}^\infty\e^{-z^2} H_{2m}(z)H_{2n+1}(|z|)\,dz}_{=: I_{mn}} \nonumber
\end{align}
In order to solve the integral $I_{mn}$, we explicitly write out the product of the Hermite polynomials,\footnote{using $H_n(z)=n! \sum_{m=0}^{\lfloor \tfrac{n}{2} \rfloor} \frac{(-1)^m (2z)^{n - 2m}}{m!(n - 2m)!}$}
\begin{align}
&H_{2n+1}(z)H_{2m}(z) =  \nonumber \\
& \; \sum_{k=0}^{n} \sum_{l=0}^{m} \frac{(-1)^{n+m-l-k}(2m)!(2n+1)!}{(2k+1)!(2l)!(n-k)!(m-l)!}  (2x )^{2k+2l+1}, \nonumber
\end{align}
and interchange sum and integral.
Then, we use
$
 \int_0^\infty(2x)^{2k+2l+1}\e^{-x^2}\,dx=2^{2k+2l+1} (k+l)! /2. \nonumber
$
Putting everything together, we obtain for $I_{mn}$ after some tedious algebra 
\begin{align}
 I_{mn}=&\sum_{k=0}^{n} \sum_{l=0}^{m} \frac{(-1)^{n+m-l-k}(2m)!(2n+1)!2^{2k+2l}}{(2k+1)!(2l)!(n-k)!(m-l)!}(k+l)!		\nonumber\\
 =&\frac{(-2) ^{m+n}(2m-1)!!(2n+1)!!}{(2n+1-2m)} \label{eq:int}
\end{align}
Inserting the normalization factors $b_n$ from Eq.\ \eqref{psin}, the overlap coefficients are given by:
\begin{equation}
c_{mn}=\sqrt[]{\frac 2 \pi}(-1)^{m+n} \frac{(2m-1)!!(2n+1)!!}{\sqrt[]{(2m)!(2n+1)!}(2n+1-2m)},
\label{cmn}
\end{equation}
and the time-dependent wave function reads
\begin{align}
&\Psi_m(z,t)
=\underbrace{\frac{\sqrt[]{2}}{\pi^{\frac 3 4}}\e^{-\frac i 2 t-\frac{z^2}{2}}\frac{(-1)^m(2m-1)!!}{\sqrt[]{(2m)!}}	}_{=: f_m(z,t)}		\nonumber\\
&\cdot \sum_{n=0}^\infty \(\frac{\e^{-it}}{\sqrt[]{2}}\)^{2n+1} \frac{(-1)^n(2n-1)!!H_{2n+1}(|z|)}{(2n)!(2n+1-2m)},
\label{Psim}
\end{align}
where we have introduced the function $f_m(z,t)$ for convenience, since it will appear again later on.

Some general remarks on the wave functions $\Psi_m(z,t)$ are in order:

(i) The energy expectation value $\langle \psi_{2m} | \hat{H}_{\mathrm{rel}}^{\kappa=\infty} | \psi_{2m} \rangle= \sum_{n=0}^\infty c_{mn}^2\epsilon_{2n}$ is divergent.
This follows from the fact that the $c_{mn}^2$ show an asymptotic decay proportional to $n^{-3/2}$, which we prove in the following.
The factors $r_n:={(2n+1)!!^2}/{(2n+1)!}$, which are part of $c^2_{mn}$, obey the recurrence relation for large $n$
\begin{align}
\frac{r_{n+1}}{r_n}
=\frac{1+\frac{3}{2n}}{1+\frac{1}{n}} \approx 1+\frac{1}{2n} \approx \sqrt[]{\frac{n+1}{n}} \label{taylor}
\end{align}
and thus $r_n$ grows  as $\propto \sqrt{n}$ for large values of $n$.
On the other hand, the square of the remaining factor in Eq.\ \eqref{cmn}, namely ${1}/{(2n+1-2m)^2}$, falls off as $n^{-2}$ for large $n$, and in total, we get the $n^{-3/2}$ scaling.
Including the scaling of the energy $\epsilon_{2n}$, which grows linearly in $n$, the energy expectation value diverges.
This can be understood from a physical point of view:
None of the initial states $\psi_{2m}(z)$ vanishes at $z=0$, so they all experience the infinite repulsive interaction  of the delta-potential in the TG-regime which causes them to immediately gain infinite energy following the quench.
Since the energy expectation value is conserved during time-evolution and all the TG-eigenstates $\phi_{n}(z)$ and importantly $\phi_{2n}(z)$ vanish at $z=0$ (they have no interaction energy), the infinite energy consists of kinetic and potential energy only.
(ii) The time evolution of the initial state $\psi_{2m}(z)$ is mainly determined by those TG eigenstates, which minimize the denominator of the series in Eq.\ \eqref{Psim}, i.e.\  $\phi_{2n-2}(z)$ and $\phi_{2n}(z)$.
Obviously, the evolution of the initial ground state $| \psi_0 \rangle$ has only one main contribution which is the TG ground-state $\phi_0(z)$ and therefore behaves differently from the excited states.

(iii) Knowing the time evolution of the eigenstates, the time evolution of arbitrary wave functions can be determined by expanding it into a superposition of these eigenstates.

\begin{figure}
\centering
\includegraphics[width=\columnwidth]{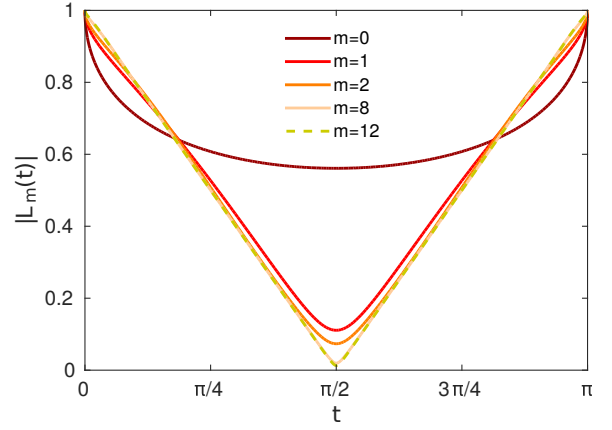}
\caption{
Fidelity between the initial state $|\Psi_m(0)\rangle=|\psi_m\rangle$ and its temporal evolution $|\Psi_m(t)\rangle$, $|L_m(t)|=\langle \Psi_m(0) | \Psi_m(t) \rangle$ for $m=\{0,1,2,8,12\}$.
\label{fig:Loschmidt}}
\end{figure}

Next, the non-trivial dynamics of the wave function can be analyzed by measuring the overlap between the initial state and the evolved state:
\begin{align} 
&	L_{m}(t):=\langle \psi_{2m}|\Psi_{m}(t)\rangle=\sum_{n=0}^\infty  \mathrm{e}^{-i \left( {2n+\frac{3}{2}} \right) t} c_{mn}^2		\nonumber\\
=&\frac{2}{\pi}\e^{-\frac 3 2 it} \frac{(2m-1)!!^2}{(2m)!}	\sum_{n=0}^\infty \left(\e^{-it}\right)^{2n} \frac{(2n+1)!!(2n-1)!!}{(2n)!(2n+1-2m)^2}, \nonumber
\end{align}
where we have inserted the coefficients from  Eq.\ \eqref{cmn}.
Having a closer look at each summand, one can show via complete induction that  
\begin{align}
\frac{(2n+1)!!(2n-1)!!}{(2n)!4(n+\frac 1 2-m)^2} =\frac{\Gamma \left( \frac 3 2+n \right)}{\Gamma \left(\frac 3 2 \right) n!} \frac{\Gamma^2 \left( \frac 1 2-m+n \right) }{4\Gamma^2 \left( \frac 3 2-m+n \right) }. \nonumber
\end{align}
This allows us to express the series $L_m(t)$ in terms of the generalized hypergeometric functions\footnote{${}_pF_q(a_1,\ldots,a_p;b_1,\ldots,b_q;z) := \sum_{n=0}^\infty \frac{(a_1)_n\cdots(a_p)_n}{(b_1)_n\cdots(b_q)_n} \, \frac {z^n} {n!}$, with the Pochhammer symbols $(x)_n:=\Gamma(x+n)/\Gamma(x)$}, which reads
\begin{align}
L_{m}(t)=  &\frac{2}{\pi}\e^{-\frac 3 2i t} \frac{(2m-1)!!^2}{(2m)!(2m-1)^2} \nonumber\\
&\;{}_3F_2\left(\frac 3 2, \frac 1 2-m, \frac 1 2-m; \frac 3 2 -m, \frac 3 2 -m, \e^{-2it}\right). \nonumber
\end{align}
This expression can be simplified for $m=0$ by using the Maclaurin series of the inverse sine.
We get
\begin{align}
L_0(t) &=\e^{-\frac i 2 t}\frac{2}{\pi}\sum_{n=0}^\infty  \frac{1}{2^{2n}}\frac{(2n)!}{(n!)^2 } \frac{\left(\e^{-it}\right)^{2n+1}}{2n+1} \nonumber \\
&=\e^{-\frac i 2 t}\frac{2}{\pi}\arcsin(\e^{-it}).
\label{arcsin}
\end{align}
In figure \ref{fig:Loschmidt}, $|L_m(t)|$ (i.e.\ the fidelity $F$) is shown.
We observe a clear difference between the evolution of the ground state fidelity and that for excited states.
This difference has its origin in the comment provided above, see (ii).
All excited states have almost vanishing overlap for a half oscillation period, $t=\pi/2$, and therefore are almost orthogonal to the initial state, in sharp contrast to the ground-state.
Other than that, the overlaps do not show any additional changes for higher $m$ (the lines for $m=8$ and $m=12$ lie almost on top of each other).
With rising $m$, we observe that the overlaps approach a linear behavior as one can see from the Fig.\ \ref{fig:Loschmidt}, due to the hypergeometric function ${}_3F_2$.

\subsection{Analysis of the ground state evolution}

We now take a closer look at the time evolution of the ground-state $\Psi_0(z,t)$.
For $m=0$, the sum in Eq.\ \eqref{Psim} can be evaluated explicitly (see appendix \ref{app:Derivation}), yielding  
\begin{align}
\Psi_{0}(z,t)=\pi^{-\frac{1}{4}}\e^{-\frac{i}{2}t}\e^{-\frac{z^2}{2}}\mathrm{erf}\left(|z|g(t)\right).
\label{Psi0}
\end{align}
with $g(t)={e^{-i t}}/{\sqrt{1-e^{-2i t}}}$.
This wave function has the obvious spatial inversion symmetry $\Psi_0(z,t)=\Psi_0(-z,t)$ as well as the  temporal periodicity $\Psi_0(z,t)=\Psi_0(z,t+\pi)$ and is symmetric around the time instant $t=\pi/2$, $|\Psi_0(z,\pi/2+t)|^2=|\Psi_0(z,\pi/2-t)|^2$.\footnote{The latter can be seen more easily from Eq.\ \eqref{Psim}. We remark that these properties are shared by all $\Psi_m(z,t)$.}
Since the wave function $\Psi_{0}(z,t)$ represents an infinite superposition of functions that vanish at the origin [see Eq.\ \eqref{Psim} for $m=0$], however, the initial wave function $\psi_0$ is finite at $z=0$, attention must be paid performing the limit $z \to 0$.
Taking the limit $t\to0$ first, however, leads to the correct initial condition, $\psi_0$.
A further  discussion of the limits $z \rightarrow 0$ and $t \rightarrow 0$ is given in the appendix \ref{app:Derivation}.
\begin{figure}[h]
\centering
\includegraphics[width=1\columnwidth]{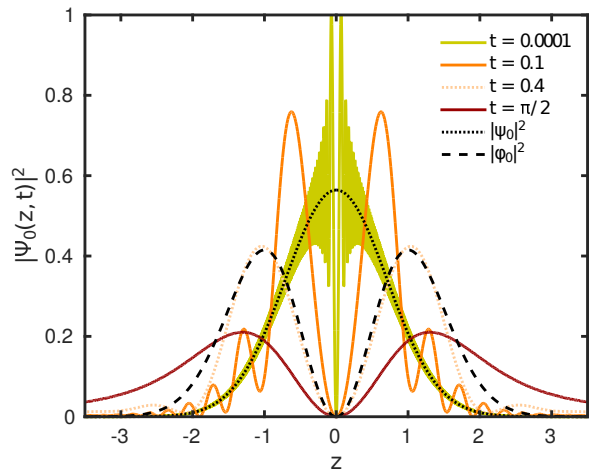}
\caption{Probability density of the temporal evolution of the ground state $|\Psi_0(z,t)|^2$ at the time instants $t=\{0.0001,0.1,0.4,\pi/2\}$.
Also plotted are the bosonic and symmetrized fermionic ground state densities $|\psi_0|^2$ and $|\phi_0|^2$, respectively.}
\label{fig:wfkt}
\end{figure}
The time evolution of the probability density is shown in figure \ref{fig:wfkt} and gives more insight in these characteristics:
At $t=0.0001$ and for relative distances $|z|>1$, the density almost perfectly coincides with the initial density $|\psi_0(z)|^2$, corresponding to the bosonic ground state.
However, close to the origin, heavy oscillations can be observed due to the fact that at $z = 0$ the wave function  $\Psi_m(z,t)$ does not converge to the initial condition (for all times $t$).
The reason herefore is that we have quenched the boundary conditions for the Hamiltonian at $z=0$. 
This is, to some extent, reminiscent of the Gibbs phenomenon in the context of Fourier transformations.
The oscillatory behavior moves closer to the origin and becomes more pronounced for smaller times $t$.
For larger times, $t<\pi/2$, the amplitudes and frequencies of the oscillations become smaller until only two smooth peaks are left over, similar to the density of a fermionic ground state.
In summary, the overall density performs a crossover between fermionic and  bosonic characteristics.

Before closer analyzing the dynamics following the quench, we mention two issues.
First, the fact that the energy expectation value diverges is reflected by the wave function's asymptotic behavior.
Looking at the asymptotic behavior of the error function for large values of $z$ 
\begin{align}
\erf(g(t)z)	\approx		-\frac{\e^{-\left(	g^2(t)\right)z^2}}{\sqrt[]{\pi g^2(t)}}\(\frac 1 z +\mathcal{O}\(\frac {1} {z^3}\)\), \label{asy}
\end{align}
with $g^2(t)=-\frac{1}{2} -\frac{i}{2}\cot{t}$.
Therefore, Eq.\ \eqref{Psi0} leads to  the asymptotic decay  $|\Psi_0(z,t)| \approx 1/|z|$.
Secondly, calculating the overlap of the initial state $\psi_0(z)$ with its evolution [see Eq.\ \eqref{Psi0}], one arrives at Eq.\ \eqref{arcsin} again.

Next, we determine the temporal evolution of the the SDM [see Eq.\ \eqref{SDM}] and its momentum distribution [see Eq.\ \eqref{nvonk}] in order to distinguish between fermionic and symmetrized fermionic attributes, which cannot be revealed by the density distribution. 
Both quantities are derived for the two-particle wave function $\Xi(z_1,z_2,t)$, where we choose the CM-state as the ground-state $\chi_0(Z)$.
Then, the two-particle wave function reads $\Xi(z_1,z_2,t)=\Psi_0(z,t) \chi_0(Z)\e^{-{\frac i 2 t}}$, where the trivial phase comes from the temporal evolution due to the CM-Hamiltonian.
The multiplication of $\Psi_0(z,t)$ with the CM state $\chi_0(Z)$ and integration over the second particle leads to smooth functions for  one-particle quantities such as the SDM as well as its momentum distribution.
Especially, the heavy oscillations occurring in $\Psi_0(t)$ are smeared out.

\begin{figure*}
\centering
\includegraphics[width=\textwidth]{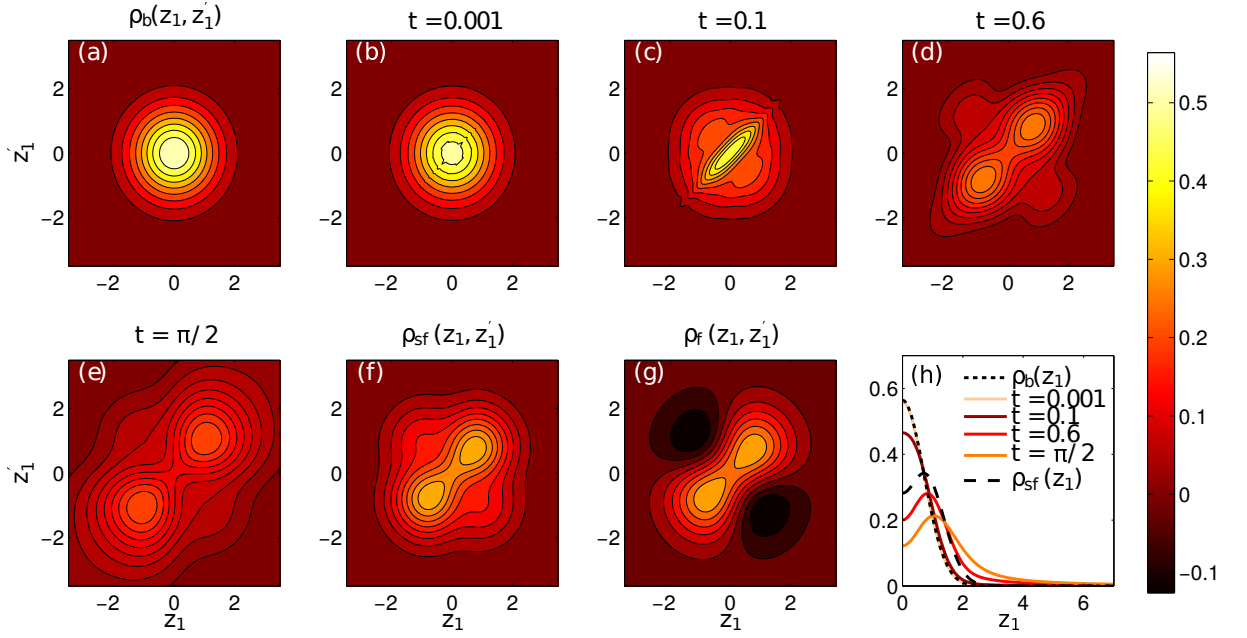}
\caption{Panels (b)-(f) show the SDM, $\rho(z_1,z_1',t)$, at different instants in time $t=\{0.001,0.1,0.6,\pi/2\}$.
Panels (a), (f) and (g) show  the ground-states SDMs for $\rspdm_b(z_1,z_1')$, $\rspdm_f(z_1,z_1')$ and $\rspdm_{sf}(z_1,z_1')$, respectively.
The single-particle densities in the laboratory frame are shown in panel (h), with $\rspdm_f(z_1)=\rspdm_{sf}(z_1)$.}
\label{fig:SDM}
\end{figure*}

We show the SDM for different time instants in Fig.\ \ref{fig:SDM}.
For small times (see Fig.\ \ref{fig:SDM}b), we observe that the SDM recovers the distribution of the initial state, namely the bosonic SDM $\rho_b(z_1,z_1')$ (Fig.\ \ref{fig:SDM}a).
At later times (Fig.\ \ref{fig:SDM}c-e), the circular symmetry is broken due to the repulsion of the bosons, until the SDM at $t=\pi/2$ becomes similar to the symmetrized fermionic SDM $\rspdm_{sf}(z_1,z_1')$ (Fig.\ \ref{fig:SDM}f).
We can here see that the temporal evolution of the wave function exhibits symmetrized fermionic rather than fermionic character (Fig.\ \ref{fig:SDM}g).
The diagonal of the SDM, i.e.\ the single-particle density, is given in Fig.\ \ref{fig:SDM}h.
It shows the crossover from a bosonic density distribution (one-centered peak) to a (symmetrized) fermionic density distribution (two separated peaks) at $t=\pi/2$, however  $\rho_{sf}(z_1)$ is stronger peaked and narrower than $\rho(z_1,z_1,t)$ at $t=\pi/2$.
After $t=\pi/2$, the density as well as the SDM develop back into the initial state and the process is repeated periodically.

\begin{figure}
\centering
\includegraphics[width=\columnwidth]{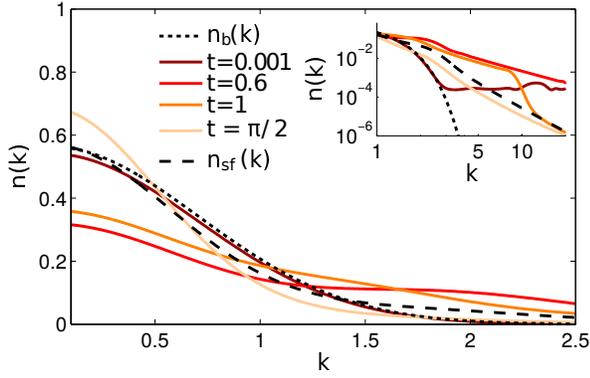}
\caption{Momentum distribution of the time evolution of the ground state at different instants in time  $t=\{0.001,0.6,1,\pi/2\}$, as well as the momentum distribution of the ground states $n_b(k),n_f(k)$ and $n_{sf}(k)$, respectively.
The inset shows the momentum distribution in logarithmic scale.
}
\label{fig:mom}
\end{figure}
The time-evolution of the momentum distribution (see Fig.\ \ref{fig:mom}) shows a crossover between the bosonic distribution $n_b(k)$ and a momentum distribution similar to $n_{sf}(k)$.
Considering half an oscillation period, the following more detailed picture emerges.
For short times, the momentum distribution is very close to the bosonic one, $n_b(k)$.
At later times, the momentum distribution first develops an extended tail for larger momenta, and then, it approaches a shape, similar to $n_{sf}$, however, with a higher and sharper peak at the origin and a more rapid decay for large $k$ values.
After $t=\pi/2$ the temporal evolution of the momentum distribution is reverted until the full period is elapsed.
For large momenta, we observe that the momentum distribution falls off first proportional to $k^{-2}$ and then the decay changes to $k^{-4}$ (see the inset of Fig.\ \ref{fig:mom}).

\section{Interaction quench from infinity to zero} \label{sec:fromInfinity}
%====================================

Let us now perform the above analysis for a reverse interaction quench, namely for $\kappa=\infty\to0$ with an eigenstate of the TG regime $\phi_{2m}(z)$ as the initial state.
We first derive the analytical formulas for the corresponding solution of the time-dependent Schr\"odinger equation and the fidelity.
Second, we deduce a closed expression for the temporal evolution of the ground state as the initial condition.
We end with a discussion of the dynamics of the SDM and the momentum distribution of the ground state evolution.

\subsection{Quench of arbitrary eigenstates}
%-----------------------------------------

We have to solve the Schr\"odinger equation $\hat{H}^{\kappa=0}_{\mathrm{rel}}|\Phi_m(t)\rangle=i\partial_t|\Phi_m(t)\rangle$ for the initial condition $\Phi_m(z,t=0)=\phi_{2m}(z)$. 
The computation of the overlap integrals, $c_{nm}=\langle \phi_{2m}|\psi_{2n}\rangle$ is similar to Eq.\ \eqref{cmn}, it is only the order of the indices which is interchanged.
This in turn implies a different decay behavior for the coefficients $c_{nm}$.
In a similar fashion to Eq.\ \eqref{taylor}, we find that the squares of the coefficients fall off with $n^{-\frac{5}{2}}$, indicating a converging, i.e.\ finite energy expectation value.
For the time evolution of the wave function $\Phi_m$, we find [for the definition of $f_m(z,t)$, see Eq.\ \eqref{Psim}]:
\begin{align}
 &\Phi_m(z,t)=
 f_m(z,t)\sqrt{2m+1}		\nonumber\\
 &\cdot \underbrace{\sum_{n=0}^\infty \(\frac{\e^{-it}}{\sqrt{2}}\)^{2n}\frac{(-1)^n(2n-1)!!H_{2n}(z)}{(2n)!(2m+1-2n)}}_{=:\Sigma_m(z,t)},
 \label{Phim}
\end{align}
where we have introduced the abbreviation $\Sigma_m$ for the occurring series.

\begin{figure}
\centering
\includegraphics[width=1\columnwidth]{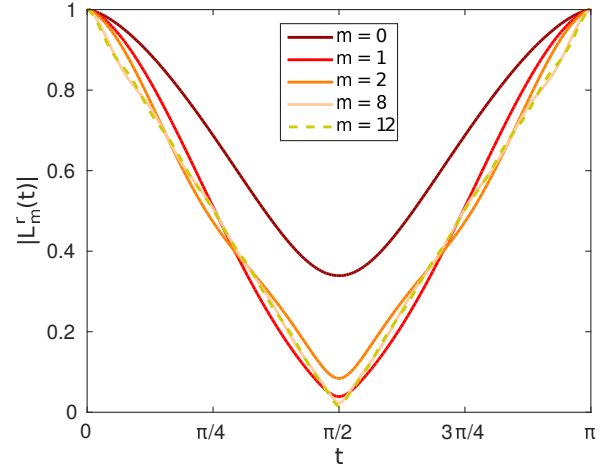}
\caption{
Fidelity $|L_m^r(t)|$ between the initial state $|\phi_{2m}\rangle$ and $|\Phi_m(t)\rangle$ for various $m=\{0,1,2,8,12\}$.}
\label{fig:Loschmidtrev}
\end{figure}

Knowing $\Phi_m(z,t)$, the overlap $L^r_m(t):=\langle\phi_{2m}|\Phi_{m}(t)\rangle$ between the initial state and the corresponding solution of the time-dependent Schr\"odinger equation can be calculated.
The superscript $r$ is used to indicate the reverse quench
\begin{align}
L^r_{m}(t)=&\frac{2}{\pi}\frac{(2m+1)!!^2}{(2m+1)!}\sum_{n=0}^\infty\frac{(2n-1)!!^2}{(2n)!(2m+1-2n)^2}\left(\e^{-it}\right)^{2n+\frac 1 2} \nonumber\\
=&\e^{-\frac i 2 t}\frac{2}{\pi}\frac{(2m+1)!!^2}{(2m+1)!} \nonumber\\
&\cdot\frac{\, _3F_2\left(\frac{1}{2},-\frac 1 2 -m,-\frac 1 2-m;\frac{1}{2}-m,\frac{1}{2}-m;e^{-2it}\right)}{(2 m+1)^2}. \nonumber
\end{align}
This expression can be simplified for $m=0$, and we find  [in analogy to the derivation of Eq.\ \eqref{arcsin}],
\begin{equation}
L^r_0(t)=\e^{-\frac i 2 t}\frac 2 \pi \(\sqrt{1-\e^{-2 i t}}+\e^{-i t} \arcsin \left(\e^{-i t}\right)\)
\end{equation}
In Fig.\ \ref{fig:Loschmidtrev}, the absolute values of some of the first few overlaps $L^r_m(t)$ are shown ($m=0,1,2,8,12$).
We again observe that all excited states have almost vanishing overlap at $t=\pi/2$, and only the ground state behaves differently.
For higher $m$, the fidelities approach a triangular shape.
In that sense, the fidelities for the reverse quench behave very similarly to the one discussed in section \ref{sec:toInfinity} for the quench from zero to infinity (c.f. Fig.\ \ref{fig:Loschmidt}).

The quantum dynamics of the two different quench scenarios are quite similar, and indeed, there is a mathematical connection between the two time-dependent wave functions $\Psi_m(z,t)$ [see Eq.\ \eqref{Psim}] and $\Phi_m(z,t)$ [see Eq.\ \eqref{Phim}, especially the definition of $\Sigma_m$(z,t)], namely
\begin{align}
\Psi_m(z,t)=\frac{\e^{it}}{\sqrt[]{2}}f_m(z,t)\cdot\(\frac{d}{dz}\Sigma_m\)(|z|,t).	\label{FromPsitoPhi}
\end{align}
which has to be compared with Eq.\ \eqref{Phim} for $\Phi_m(z,t)$.
A derivation of this equation Eq.\ \eqref{FromPsitoPhi} is given in the appendix \ref{app:relation}.
We see that apart from the common factor $f_m(z,t)$, $\Psi_m$ and $\Phi_m$ are connected via integration (differentiation).
Hence, $\Psi_m$ and $\Phi_m$ are connected in a similar way as $\psi_{2m}$ and $\phi_{2m}$\footnote{See the recursion relation for the Hermite polynomials.}.
With the help of this relation, we derive a closed expression for the time evolution of the ground state $\Phi_0(z,t)$ in the next subsection.

\subsection{Time evolution of the ground state}
%---------------------------------------------------------

\begin{figure}
\centering
\includegraphics[width=\columnwidth]{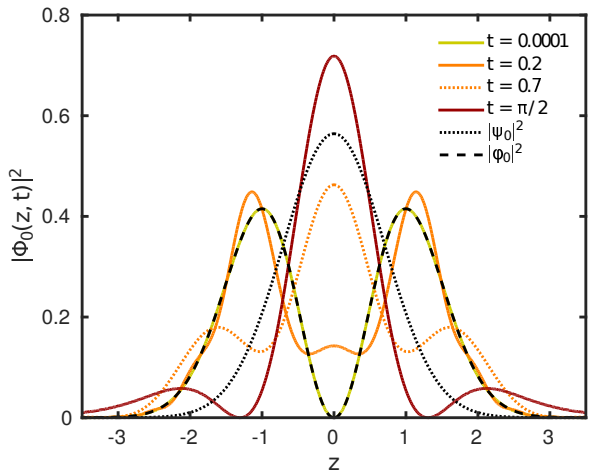}
\caption{Probability density of the time evolution of the ground state $|\Phi_0(z,t)|^2$ at time instants $t=\{0.0001,0.2,0.7,\pi/2\}$.
Also shown are the bosonic and symmetrized fermionic ground state densities $|\psi_0(z)|^2$ and $|\phi_0(z)|^2$, respectively.}
\label{fig:wfktrev}
\end{figure}
We can find a closed form for $\Phi_0(z,t)$ by simply inverting Eq.\ \eqref{FromPsitoPhi} and using the closed form of $\Psi_0(z,t)$ [see Eq.\ \eqref{Psi0}].
The emerging integral of the error function can be calculated via integration by parts and we obtain
\begin{equation}
\Phi_0(z,t)=\frac{\sqrt[]{2}}{\pi^{\frac 1 4}}		\e^{-\frac 3 2 it}\e^{-\frac{z^2}{2}} \(\frac{1}{\sqrt[]{\pi}g(t)}\e^{-z^2g^2(t)}     +z\, \erf\(	z g(t)		\)\)\label{Phi0}
\end{equation}
with $g(t)={e^{-it}}/{\sqrt[]{1-\e^{-2it}}}$  from Eq.\ \eqref{Psi0}.
$\Phi_0(z,t)$ possesses the same symmetries in space and time as $\Psi_0(z,t)$.
Using Eq.\ \eqref{asy}, we obtain that $|\Phi_0(z,t)|$ has an asymptotic decay proportional to $|z|^{-2}$.

A crossover between the (symmetrized) fermionic and a Gaussian shaped density distribution can be seen, which is stronger peaked at the center and exhibits two wing peaks  in comparison to the bosonic density $|\psi_0|^2$.
In contrast to the probability density of $\Psi_0(z,t)$ (cf. Fig.\ \ref{fig:wfkt}), we do not observe any fast oscillations around the origin since now, the initial condition can be expressed by the eigenstates of the Hamiltonian in a uniformly converging manner.
The peculiar behavior that have occurred due to the error function in $ \Psi_0(z,t)$ [cf. Eq.\ \eqref{Psi0}] is now absent due to the additional factor $z$ [see last term in Eq.\ \eqref{Phi0}].
Since at  $t=0$, the exponential term $\e^{-z^2g^2(t)}$ vanishes and the error function approaches the sign-function, the wave function properly accounts for the initial condition.

\begin{figure*}
 \centering
 \includegraphics[width=\textwidth]{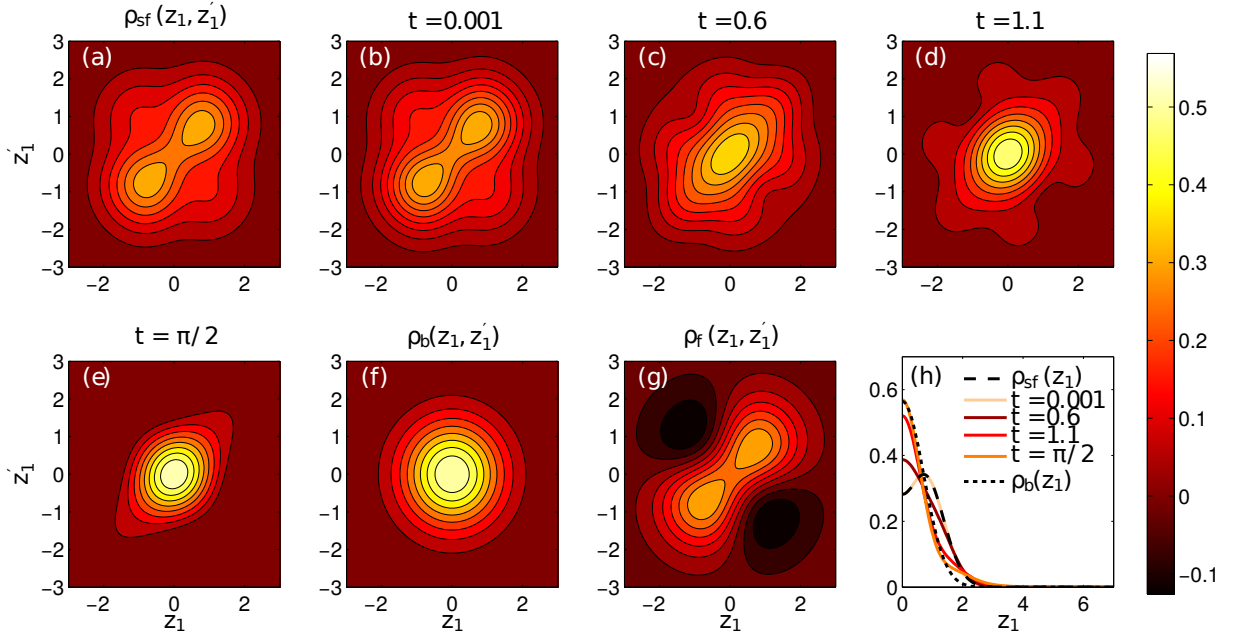}
 \caption{The time evolution of the SDM for the initial ground state at different instants in time $t=\{0.001,0.6,1.1,\pi/2\}$, as well as the ground state SDMs $\rspdm_b(z_1,z_1')$ (bosonic), $\rspdm_f(z_1,z_1')$ (fermionic) and $\rspdm_{sf}(z_1,z_1')$ (symmetrized fermionic).
 The densities of these states are shown in (h), with $\rspdm_f(z_1)=\rspdm_{sf}(z_1)$.}
 \label{fig:RSPDMrev}
 \end{figure*}

Using the framework described in section \ref{sec:Stationary}, the SDM is obtained numerically [by using Eq.\ \eqref{Phi0} and choosing the ground-state $\chi_0(Z)$ for the CM-coordinate] and is shown in Fig.\ \ref{fig:RSPDMrev} for different instants in time.
We observe that the initial symmetrized fermionic distribution attains an increasing elliptical symmetry at later times $t$.
The two initial maxima merge with each other and form a new, single maximum, centered at the origin leading to a state similar to the bosonic one $\rspdm_b(z_1,z_1')$ (see Fig.\ \ref{fig:RSPDMrev}f). 
The single-particle density is given in Fig.\ \ref{fig:RSPDMrev}h and shows  a transition between the shapes of the symmetrized fermionic and the bosonic density.

\begin{figure}
\centering
\includegraphics[width=\columnwidth]{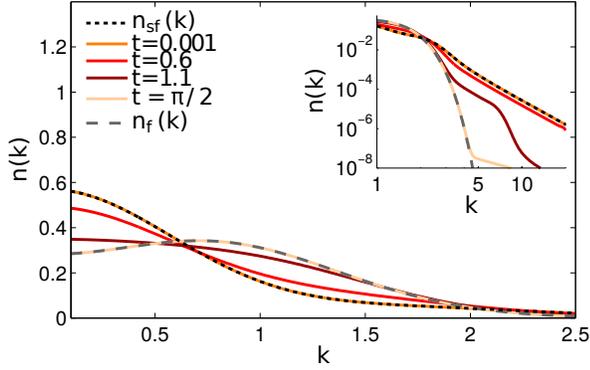}
\caption{Time evolution of the momentum distribution at different instants in time  $t=\{0.001,0.6,1.1,\pi/2\}$, as well as the momentum distribution of the ground states $n_b(k),n_f(k)$ and $n_{sf}(k)$, respectively.
The inset shows the momentum distribution in logarithmic scale.}
\label{fig:momrev}
\end{figure}

The momentum distribution on the other hand, behaves quite differently (see Fig.\ \ref{fig:momrev}).
We identify a transition between the initial state $n_{sf}(k)$ and the fermionic momentum distribution $n_f(k)$.
The maximum value at the origin slowly decreases and a new maximum at a higher value of $k$ emerges.

\section{Multiple interaction quenches} \label{sec:double}
%=========================================================

\begin{figure}
\centering
\hspace{5mm}

\includegraphics[width=0.8\columnwidth]{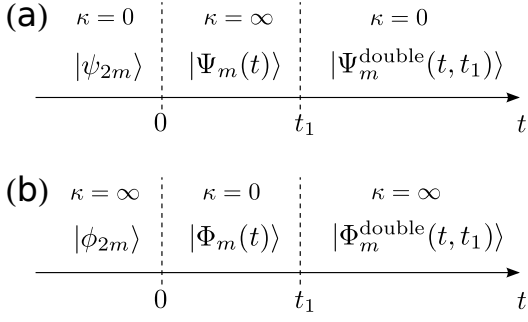}
\caption{Double quench scenarios as discussed in the main text.}
\label{fig:sketch}
\end{figure}

In this section, we consider double interaction quenches:
In the first scenario (a), the initial wave function $\psi_{2m}$ is quenched from $\kappa=0 \to \infty$ at $t=t_0=0$, yielding $\Psi_m$, and after some time $t=t_1$  a quench back to the non-interacting regime takes place.
The corresponding solution of the time-dependent Schr\"odinger equation shall be denoted as $\Psi^{\text{double}}_m$.
In the second scenario (b), we start with the initial state $\phi_{2m}$, an eigenstate in the TG-regime, and quench it to the non-interacting regime, yielding $\Phi_m$.
At time $t=t_1$, the reverse quench $\kappa=0 \to \infty$ is applied and the dynamics is governed by the corresponding solution of the time-dependent Schr\"odinger equation being labeled by $\Phi^{\text{double}}_m$.
These scenarios are sketched in figure \ref{fig:sketch}.

We start with the double-quench scenario (a).
In order to calculate the time evolution  $| \Psi^{\text{double}}_m(t,t_1) \rangle$, we insert Eq.\ \eqref{Psim} into the general time-evolution of $| \Psi^{\text{double}}_m(t,t_1) \rangle$ [c.f. Eq.\ \eqref{howtoevolve}].
\begin{align}
& | \psidouble_m(t,t_1) \rangle \nonumber\\
=&\sum_{k=0}^\infty  \e^{-iE_{2k}(t-t_1)}|\psi_{2k}\rangle	\langle \psi_{2m} | \Psi_m(t_1) \rangle \label{psi_d} \\
=&\sum_{k=0}^\infty  \e^{-iE_{2k}(t-t_1)}|\psi_{2k}\rangle		\sum_{n=0}^\infty \e^{-iE_{2n+1}t_1}\langle \psi_{2k}	 |\phi_{2n}\rangle c_{mn} 
 \nonumber \\
=&
\frac{2(-1)^m}{\pi}	\frac{(2m-1)!!	}{\sqrt{(2m)!}}	\sum_{k=0}^\infty	(-1)^k\e^{-iE_{2k}(t-t_1)} \frac{(2k-1)!!}{\sqrt{(2k)!}}|\psi_{2k}\rangle	\nonumber\\ 
&\cdot\e^{-\frac 3 2 i t_1}\frac{\, _3F_2\left(\frac{3}{2},\frac{1}{2}-k,\frac 1 2 -m;\frac{3}{2}-k, \frac  3 2-m;\e^{-2 i t_1}\right)}{(2 k-1)(2m-1)} \nonumber
\end{align}
where we have expressed the sum over $k$ in terms of the generalized hypergeometric functions.
When looking at the absolute value of the overlap between the initial state and the corresponding solution of the time-dependent Schr\"odinger equation for time instants $t>t_1$, $| \langle \psi_{2m}|\psidouble_m(t,t_1) \rangle |$, we find that it only depends on the time instant $t_1$ at which we perform the second quench back to the initial non-interacting regime, and not on the actual time $t$.
This quite general feature can be seen directly by projecting  Eq.\ \eqref{psi_d} onto $| \psi_{2m} \rangle$ and using $\langle\psi_{2m}|\psi_{2k}\rangle=\delta_{m,k}$
\begin{align}
&\left| \langle\psi_{2m}|\psidouble_m(t,t_1)\rangle	\right| =\left| \sum_{k=0}^\infty  \e^{-iE_{2k}(t-t_1)} \delta_{k,m}		\langle \psi_{2k}	|\Psi_{m}(t_1)\rangle  \right| \nonumber \\
&=\left| \langle \psi_{2m}	|\Psi_{m}(t_1)\rangle \right| =| L_m(t_1) | \nonumber
\end{align}
In particular, the absolute values of the overlap is the same as for the $\kappa: 0 \to \infty$ quench (see section \ref{sec:toInfinity}), except that it does not depend on $t$ but exclusively on $t_1$.
Thus, once the second quench is conducted, the absolute value of the overlap remains constant.

Focusing on scenario (b), a similar procedure can be applied to express $\Phi^{\text{double}}_m$ analytically:
\begin{align}
&|\Phi_m^{\text{double}}(t,t_1)\rangle=\sum_{k=0}^\infty \e^{-iE_{2k+1}(t-t_1)} |\phi_{2k}\rangle \sum_{n=0}^\infty		\e^{-iE_{2n}t_1} c_{nk}c_{nm}	\nonumber\\
&=\frac{2(-1)^m(2m+1)!}{\pi\sqrt[]{(2m+1)!}}			\sum_{k=0}^\infty   \e^{-iE_{2k+1}(t-t_1)}\frac{(-1)^k	(2k+1)!}{\sqrt[]{(2k+1)!}} |\phi_{2k}\rangle \nonumber\\
&\cdot  \e^{-\frac i 2 t_1}\frac{\, _3F_2\left(\frac{1}{2},-\frac 1 2-m,-\frac 1 2-k;\frac{1}{2}-m,\frac{1}{2}-k;e^{-2it_1}\right)}{(2k+1)(2 m+1)} \nonumber
\end{align}
and we get the constant absolute value of the overlap
\begin{equation}
| \langle\phi_{2m}|\Phi_m^{\text{double}} (t,t_1) \rangle | =| L_m^r(t_1)|. \nonumber
\end{equation}

\section{Conclusion} \label{sec:conclusion}
%===================

In this work, we have investigated  the quantum dynamics of two bosons following an interaction quench from both zero (non-interacting regime) to infinity (Tonks-Girardeau regime) and from infinity to zero.
The interaction potential has been modeled by a contact interaction.

We have established analytical expressions for the two-body time-dependent wave functions and fidelities (Loschmidt echoes) for both quench scenarios, taking arbitrary (bosonic) eigenfunctions of the initial Hamiltonian as initial states. Therefore, the dynamics of a general initial wave packet can be studied by investigating a corresponding superposition of these eigenfunctions' dynamics.

For the ground state, we have found closed expressions for the dynamical evolution of its wave function.
The corresponding temporal evolution of the reduced single-particle density matrix as well as the momentum distribution have been calculated on basis of this wave function.
We have observed a characteristic crossover in these quantities between bosonic, symmetrized fermionic as well as fermionic behavior.
Additionally, we have shown that in the course of the time evolution  excited states become almost orthogonal w.r.t.\ to the initial state at half an oscillation period, $t=\pi/2$, which is a behavior that the evolution of the ground state does not exhibit.

Although there are many similarities between the quench from zero to infinity and the reverse quench, there is one striking difference between them:
For the reverse quench, the energy expectation value remains finite whereas it diverges for the quench into the Tonks-Girardeau regime.

Knowing the analytical expression for single quenches, we have studied double quench scenarios:
From the non-interacting to the TG regime and then back to the non-interacting regime, and vice versa.
In doing so, we have established closed expressions for the fidelity (Loschmidt Echo).

Apart from the insight into the problem's peculiarities and the gained intuition, our solutions can be used as a time-dependent correlated two-particle basis which could serve as a starting-point for numerical calculations or perturbative treatments of more complicated bosonic many particle systems.

The here presented framework can be used to evaluate the dynamics of two bosons in three dimensions following an interaction quench, where the interaction is modeled by the regularized contact interaction assuming s-wave scattering.
The occurring three-dimensional overlap integrals can be linked to the overlap integrals evaluated in this work:
the overlap integrals resulting from an interaction quench from zero to infinity in three dimensions can be linked to the one-dimensional overlap integrals resulting from an interaction quench from infinity to zero, and vice versa.

\section*{Acknowledgements}
The authors thank Sven Kr\"onke for many fruitful discussions.
Financial support by the Deutsche Forschungsgemeinschaft (DFG) in the framework of the SFB 925 “Light induced dynamics and control of correlated quantum systems” is gratefully acknowledged by P.S.

%=========================================================
%=========================================================
%=========================================================
\appendix

\section{Derivation of the closed expression for $\Psi_0(z,t)$} \label{app:Derivation}
%==========================================================

Here, we derive the closed expression for the wave function $\Psi_0(z,t)$ given in Eq.\ \eqref{Psi0}.
Inserting $m=0$ into Eq.\ \eqref{Psim} and using the relation $H_{2n}(0)=(-1)^n 2^n (2n-1)!!$, we can write the time evolution of the wave function as
\begin{align}
\Psi_{0}(z,t)=
\e^{-\frac i 2 t } \frac{2}{\pi^{\frac 3 4}}\e^{-\frac{z^2}{2}}\sum_{n=0}^{\infty}\left(\frac{\e^{-i t}}{2}\right)^{2n+1}\frac{H_{2n+1}(|z|)H_{2n}(0)}{(2n+1)!} \label{Psi0open}
\end{align}
In order to apply Mehler's formula,
\begin{align}
\sum_{n=0}^\infty\frac{H_n(x)H_n(y)}{n!}\left(\frac{u}{2}\right)^n=\frac{1}{\sqrt{1-u^2}}\mathrm{e}^{\frac{2xyu-(x^2+y^2)u^2}{1-u^2}}\label{Mehler},
\end{align}
we transform the first Hermite polynomial by using the recurrence relation $\frac{d}{dx}H_n(x)=2nH_{n-1}(x)$ and $H_{2n+1}(0)=0$:
\begin{align}
&\frac{d}{dx}\sum_{n=0}^\infty\left(\frac{u}{2}\right)^{2n+1}\frac{H_{2n+1}(x)H_{2n}(0)}{(2n+1)!} \nonumber\\
=&2\sum_{n=0}^{\infty}\left(\frac{u}{2}\right)^{2n+1}\frac{H_{2n}(x)H_{2n}(0)}{(2n)!}\nonumber\\
=&u\sum_{n=0}^{\infty}\left(\frac{u}{2}\right)^{n}\frac{H_{n}(x)H_{n}(0)}{n!}\nonumber\\
=&u\cdot(1-u^2)^{-\frac 1 2}\e^{-x^2\frac{u^2}{1-u^2}}		\label{closing}.
\end{align}
where we have used $x=|z|$ and $u=\e^{-it}$.
Next, we integrate w.r.t.\ $x$ yielding the closed form [see Eq.\ \eqref{Psim}]:
\begin{align}
\Psi_{0}(z,t)=\pi^{-\frac{1}{4}}\e^{-\frac{i}{2}t}\e^{-\frac{z^2}{2}}\mathrm{erf}\left(|z|\frac{e^{-i t}}{\sqrt{1-e^{-2i t}}}\right)
\label{app:Psi0}
\end{align}
The above calculations have to be considered critically:
Not only do we operate on the radius of convergence of Mehler's formula (which holds for $u \in \mathbb{C}$ and $|u| < 1$), we also interchange the differentiation and summation to get from Eq.\ \eqref{closing} to Eq.\ \eqref{app:Psi0}.
We can show that the wave function \eqref{app:Psi0} fulfills the Schr\"odinger equation for $t >0$ and is  normalized to one.
Therefore, we know that the closed form \eqref{app:Psi0}  is correct if it converges to the initial condition at $t=0$.
Taking the limit $t\to0$ first,  the wave function approaches the initial Gaussian (since then, the error function approaches 1), however, taking first $z\to 0$ and then $t\to 0$, the wave function \eqref{app:Psi0} vanishes at the origin and violates the initial condition.

\section{Derivation of the relation between $\Psi_m(z,t)$ and $\Phi_m(z,t)$} \label{app:relation}
%=========================================================

In order to prove the relation between the wave functions $\Psi_m$ [see Eq.\ \eqref{Psi0}] and $\Phi_m$ [see Eq.\ \eqref{Phi0}], which is mentioned in Eq.\ \eqref{FromPsitoPhi}, we differentiate the series $\Sigma_m(z,t)$ from Eq.\ \eqref{Phim} as well as interchange differentiation and summation, leading to
\begin{align}
		&\frac{d}{dz}\sum_{n=0}^\infty \(\frac{\e^{-it}}{\sqrt[]{2}}\)^{2n}\frac{(-1)^n(2n-1)!!H_{2n}(z)}{(2n)!(2m+1-2n)}	\nonumber\\
     =&	\sum_{n=1}^\infty \(\frac{\e^{-it}}{\sqrt[]{2}}\)^{2n}\frac{(-1)^n(2n-1)!!(4n)H_{2n-1}(z)}{(2n)!(2m+1-2n)}	\nonumber\\
     =&\sqrt[]{2}\e^{-it}\sum_{n=0}^\infty \(\frac{\e^{-it}}{\sqrt[]{2}}\)^{2n+1}\frac{(-1)^{n+1}(2n-1)!!H_{2n+1}(z)}{(2n)!(2m+1-2(n+1))} \label{app:relationproof}
\end{align}
In the last step, we changed the summation index from $n$ to $n-1$.
Comparing Eq.\ \eqref{app:relationproof} to the series for $\Psi_m(z,t)$ [see Eq.\ \eqref{Psim}], we see that they are identical, except for the missing absolute value in the Hermite polynomial's argument.
Thus, Eq.\ \eqref{FromPsitoPhi} is proven.

%========================================================================
\bibliography{verweise}
\end{document}